\documentclass[reprint,amsmath,amssymb,aps,prx]{revtex4-2}
\usepackage{blindtext}
\usepackage{graphicx}
\usepackage{svg}
\usepackage{float}
\usepackage{longtable}
\usepackage{dcolumn}
\usepackage{textgreek}
\usepackage{textcomp}
\usepackage{xr}
\usepackage[acronym]{glossaries}
\usepackage{silence}
\begin{document}
\newacronym{dn}{double-network}{double-network}

\newacronym{lve}{LVE}{linear viscoelastic}
\newacronym{paam}{PAAm}{polyacrylamide}
\newacronym{laos}{LAOS}{Large Amplitude Oscillatory Shear}
\newacronym{CaCO3}{CaCO$_3$}{calcium carbonate}
\newacronym{gdl}{GDL}{Glucono delta-lactone}
\newacronym{tmao}{TMAO}{trimethylamine N-oxide}
\newacronym{gdhcl}{GdHCl}{guanidine hydrochloride}
\newacronym{gcage}{$G_{cage}$}{Cage modulus}
\newacronym{lb}{LB}{Lissajous-Bowditch}
\preprint{APS/123-QED}

\title{Mechanistic Origins of Yielding in Hybrid Double Network Hydrogels}
\author{Vinay Kopnar$^1$, Adam O'Connell$^2$, Natasha Shirshova$^3$ and Anders Aufderhorst-Roberts$^1$}
\email{vinay.h.kopnar@durham.ac.uk}

\affiliation{$^1$Department of Physics, Durham University,  Lower Mountjoy, South Rd, Durham DH1 3LE, UK}
\affiliation{$^2$Polymer Science Platform, Reckitt Benckiser Health Care UK Ltd, Dansom Lane S, Hull HU8 7DS, UK}
\affiliation{$^3$Department of Engineering, Durham University,  Lower Mountjoy, South Rd, Durham DH1 3LE, UK}

\begin{abstract}
Hybrid double-network hydrogels are a class of material that comprise transiently and permanently crosslinked polymer networks and exhibit an enhanced toughness that is believed to be governed by the yielding of the transient polymer network. The precise role of the two polymer networks in this yielding transition and their interplay remains an open question that we address here through constructing a series of hydrogel designs in which the interaction \textit{within} and \textit{between} the two polymer networks are systematically inhibited or enhanced. We characterise each of the hydrogel designs using large amplitude oscillatory shear rheology (LAOS). Inspecting yielding through elastic stress across hydrogel designs, we elucidate that the hybrid double-network hydrogel exhibits a two-step yielding behaviour that originates from the presence of transient crosslinks. Examining the rheological response within each oscillatory cycle and across the hydrogel designs, we show that the micro-structural changes in the transient network are crucial in the second stage of this yielding. We surmise that the first step of yielding is determined by the intermolecular interactions \textit{between} the two polymer networks by systematically altering the strength of the interactions. These interactions also influence the second step of yielding, which we show is governed by the transient intermolecular interactions \textit{within} the polymer networks. Our study therefore reveals that the interactions \textit{between} the polymer networks are as crucial as \textit{within} the polymer networks and therefore provides insights into how the yielding mechanisms in soft composite materials can be identified, adjusted, and controlled. 
\end{abstract}

\maketitle

\section{Introduction}

Across nature, almost all living materials are composites and comprise multiple constituent elements that have distinct and complementary mechanical properties.  Examples include spider silk, whose high tensile strength arises from the mechanical properties of different structural proteins \cite{Vierra11,denny1976physical,DeTommasi2010}, and tendons, whose stiffness and compliance strength arise from the combination of elastin and collagen \cite{Henninger2019}. Owing to the synergistic effect that combines the best characteristics of the constituents, natural composites exhibit mechanical properties that are inaccessible to single component materials \cite{burla2019mechanical}.

For this reason, composite polymeric materials present a useful model system \cite{Zhao2021,Ji2021,ElSayed2023,PLOCHER2021108669,Trask2007} for understanding the mechanical synergy exhibited in living materials \cite{Salernitano@2003}. Of particular note are hybrid \acrshort{dn} hydrogels that contain two interpenetrating polymer networks dispersed in water, one of which is transiently crosslinked through noncovalent interactions and the second of which is permanently crosslinked through covalent bonds \cite{Sun2012}. Hybrid \acrshort{dn} hydrogels undergo fracture at remarkably high strains and this has been proposed to arise from the ability of the transient crosslinks to break under deformation, thereby dissipating energy and preventing the accumulation of elastic stress within the permanently crosslinked network. Experimental \cite{Gong2010,Long2021,matsuda2016yielding} and theoretical studies \cite{Brown2007,Xiao2021,Zhu2020} have indicated that this process represents a local yielding mechanism in the hydrogel. Studies have indicated that this yielding occurs at higher applied strains in hybrid double-network hydrogels than in conventional double-network hydrogels that are entirely permanently crosslinked \cite{chen2014fracture}. However, the precise structural mechanism of this yielding remains a matter of speculation with potential mechanisms including the dissociation of polymer chains \cite{chen2014fracture}, the stretching of crosslink junctions \cite{John2019} and the breaking of crosslinks \cite{matsuda2016yielding}. In addition, the transiently crosslinked network has been proposed to reinforce the permanently crosslinked network, even after yielding \cite{Long2021}, suggesting direct intermolecular interactions between the two polymer networks \cite{Nakajima2009, Sun2012}. Despite the apparent importance of this yielding process in hybrid \acrshort{dn} hydrogels, it remains challenging to precisely identify the yielding point, to quantify the structural changes that take place during yielding, and to examine how this process relates to the intermolecular interactions between the two polymer networks. 

In this work, we address these open questions by probing the mechanical response of a model hybrid \acrshort{dn} hydrogel using oscillatory rheology. Using a well-defined existing protocol \cite{Sun2012}, we construct this hydrogel from alginate and \acrfull{paam}. The alginate/\acrshort{paam} \acrshort{dn} hydrogel is an ideal model system for a number of reasons. Firstly, its bulk mechanical properties are well-characterised \cite{baselga1987elastic,Calvet2004,Zhang2005}. Secondly, the two constituents have distinct and controllable crosslinking chemistries; specifically N,N'-methylenebis(acrylamide) permanent covalent crosslinks in the \acrshort{paam} network and the transient ionic crosslinks in the alginate network between Ca\textsuperscript{2+} and $\alpha$-L-guluronate (G) units of the alginate chains. It is also an excellent model system from the perspective of proving the effect of intermolecular interactions since the two polymers are known to form hydrogen bonds between the carboxyl group of alginate and the amide group of \acrshort{paam} \cite{Xiao2000-zy,olpan2008,Sun2012}. Using this model system therefore allows us to construct a series of hydrogel designs in which either one or both of crosslinks \textit{within} each polymer network are present. Through the use of chaotropic or cosmotropic reagents, we may also either enhance or inhibit the hydrogen bonds that govern the interaction \textit{between} the two polymer networks. 

To quantify the yielding mechanism and relate it to structural changes, we use large amplitude oscillatory shear \acrfull{laos} rheology. \acrshort{laos} is a data-rich rheological method that applies sinusoidal shear strain cycles of increasing amplitude. \acrshort{laos} reveals information about the rheological response within a given steady-state cycle while also facilitating the progressive of average changes between subsequent steady-state cycles. To study the response of a material in each deformation cycle, we decompose stress into elastic and viscous components using Fourier transform coupled methods \cite{ewoldt2008new}. Further analysis can then give insights into structural changes while allowing the identification of elastic, viscous, and yielding behavior \cite{kamkar2022large} and to visualise strain-dependant as well as strain-rate dependent phenomena. It also enables comparison of modes of yielding that a hydrogel can undergo \cite{Donley2019} and provides a way to infer microstructural changes within the hydrogel \cite{Rogers2011}. It has been used to study the mechanistic origins of the rheological behaviour of a wide variety of systems \cite{kamkar2022large} including composite hydrogels \cite{Tarashi2022,Shoaib2022,Wang2023}. However, to our knowledge, the structural and rheological impacts of varying interactions between the two polymer networks in a \acrshort{dn} hydrogel have yet to be studied.   Since \acrshort{laos} is a novel method to study a \acrshort{dn} hydrogel, it provides an attractive  approach to investigate the role of individual polymer networks on the viscoelastic transformations and yielding behavior.  

In this paper, we critically examine the hydrogel designs by utilising the \acrshort{laos} framework. We seek to: 

(1) Precisely identify and characterise the yielding transition of the hybrid double network hydrogel.  

(2) Examine how the transient crosslinks \textit{within} the  alginate network and the permanent crosslinks \textit{within} the \acrfull{paam} influence this yielding transition. 

(3) Examine the effect of interactions \textit{between} the two polymer networks on the yielding behaviour.

\section{Materials and Methods}
\subsection{Hydrogel Preparation}


All \acrshort{dn} hydrogels were prepared from sodium alginate (Sigma Aldrich, 180947) and polyacrylamide with a 0.11:0.89 mass ratio of the monomers. Initially, a calculated amount of acrylamide (AAm) (Sigma Aldrich, $\geq 99\%$) and alginate was dissolved in the deionised (DI) water to prepare solutions of AAm (5 M) and alginate (5.4 \%wt/vol). CaCO$_3$ was mixed with the required amount of DI water for a total sample volume of 5 ml and a final concentration of 0.045 M. The mixture was sonicated using the iSonic P4820 Commercial Ultrasonic Cleaner at 25 $^{\circ}$C for 20 minutes. Aliquots from alginate and AAm solutions were mixed to achieve the mass ratio 0.11:0.89. To form the reaction mixture, solutions of N, N\rq-Methylenebis(acrylamide) (MBA) (Sigma Aldrich, 99\%), a crosslinker for the AAm network, N, N, N\rq, N\rq-Tetramethylethylenediamine (TEMED) (Sigma Aldrich, 99\%), a crosslinking accelerator, and a freshly prepared solution of ammonium persulfate (APS) (Sigma Aldrich, $\geq 98\%$) were added such that the final concentration of MBA, TEMED, and APS was 0.5 mM, 2.7 mM, and 0.55 mM respectively. The reaction mixture was then purged with nitrogen for 10 seconds. Next, the suspension of sonicated CaCO\textsubscript{3} and freshly prepared solution of glucono-$\delta$-lactone (GDL) (Sigma Aldrich, $\geq 99\%$) was added. To achieve full dissociation of Ca\textsuperscript{+2} ions, the molar ratio between CaCO\textsubscript{3} to GDL was kept at 1:2. The reaction mixture was then poured into custom-designed moulds prepared for high-throughput testing (Figure S1 in Supplemental Material \cite{supp}) and placed into an oven for 3.5 hours at 50 $^{\circ}$C after sealing them with vacuum grease and plastic lid. The sealed samples were then stored at room temperature and tested after 21 hours. Unless stated otherwise, before testing, all hydrogels in this study were soaked in a solution of CaCO\textsubscript{3} and GDL with the same concentration as in the reaction mixture for 4.5 hours. The dry weight swelling ratio, defined as the ratio of the difference between swollen material weight and dried material weight to the swollen material weight was determined by drying the hydrogels in a vacuum desiccator until constant weight as determined gravimetrically and found to be 0.75±10\% i.e. the swollen weight is 4 times that of the dry weight, with the hydrogel containing 75 wt\% water. %

 
\subsection{Hydrogel Design} \label{hydrogel_design}

To illustrate the effect of individual polymer networks' rheological properties on the rheological behavior of the \acrshort{dn} hydrogel, we designed two hydrogels containing only one crosslinked network by inhibiting an \textit{intra-}polymeric crosslink systematically: Alg+/\acrshort{paam}- and Alg-/\acrshort{paam}+. Inhibiting crosslinks requires a design strategy that carefully prevents any secondary effects from arising. We synthesised ionically-crosslinked Alg+/\acrshort{paam}- hydrogel by forgoing the chemical crosslinking in the \acrshort{paam} network as illustrated in Figure \ref{figure:IntercycleLBSN}(a). Similarly, we formed Alg-/\acrshort{paam}+ hydrogel by excluding the ionic crosslinkers that make up the alginate network while keeping chemical crosslinking in the \acrshort{paam} intact as illustrated in Figure \ref{figure:IntercycleLBSN}(d). All other reaction mixture components were kept the same as described in section A.

Selectively inhibiting one form of polymer crosslinking in this way would have inherently affected the swelling of the hydrogel which in turn changes the stiffness and other mechanical properties \cite{Lee2000, Subramani2020, Chen2016}. To maintain a consistent degree of swelling between samples, all hydrogels were soaked in soaking slutions of CaCO\textsubscript{3}+GDL of the same concentration as in the reaction mixture, until the dry swelling ratio, reached 0.75. Measuring the dry swelling ratio for independent samples shows that different hydrogels reach this value over different times, Alg-/\acrshort{paam}+ hydrogel after 1.5 hours, Alg+/\acrshort{paam}- hydrogel hydrogels after 2 hours and the double-network (Alg+/\acrshort{paam}+) hydrogel after 4.5 hours.
  
To tune \textit{inter-}polymeric crosslinking, urea, \acrfull{gdhcl}, and \acrfull{tmao} with varying concentrations were introduced individually in the soaking solution and the soaking time was tuned to achieve the same swelling ratio (0.75) as the Alg+/\acrshort{paam}+ hydrogel. For urea and \acrshort{gdhcl}, the soaking time was 4.5 hours while for \acrshort{tmao}, it was 1.5 hours.

\subsection{Rheology}

We used \acrshort{laos} to study the linear and non-linear properties of the hydrogel designs. We applied oscillations from 0.1\% to 500\% strain amplitude ($\gamma_{0}$) at 0.5 Hz frequency using Netzsch Kinexus Pro stress-controlled rheometer.  We monitored how the rheological response changes within a single oscillatory cycle. In \acrshort{laos} terminology, this is often referred to as intra-cycle behavior \cite{ewoldt2008new}. We also examined inter-cycle behavior which pertains to how behavior changes over the subsequent oscillatory cycles of increasing amplitude. At each amplitude, 10 oscillatory cycles were applied which is necessary for the strain to reach a steady state strain amplitude, and the response from the final oscillatory cycle was used for analysis. Samples to test were sheared using a 40 mm sandblasted parallel plate unless otherwise stated.  A 5 N constant normal force was applied during testing to minimise sample slip. Alg+/\acrshort{paam}- hydrogel, and the samples reported later where inter-network crosslinks are strengthened were seen to be prone to slippage. In the case of Alg+/\acrshort{paam}- hydrogel, an adhesive sandpaper was attached to the top rheometer plate and a thin layer of super glue was used to adhere the top surface of the sample to the sandpaper. We checked that the use of superglue does not influence sample viscoelasticity by confirming that the frequency spectra of samples with and without superglue overlay within one standard deviation (Figure S2 in Supplemental Material \cite{supp}). For the samples with strengthened inter-network crosslinks, serrated parallel plates were used with a 20 N constant normal force. We used MATLAB to analyse stress-strain curves obtained from testing. To analyse intra-cycle behavior particularly, we made use of the MITlaos package \cite{mitResearch} which assists in extracting relevant parameters from the stress-strain curves.


\section{Hybrid Double-Network Hydrogels Undergo Continuous Yielding driven by Transient Crosslinking}

\begin{figure*}
\includegraphics[width=0.97\textwidth]{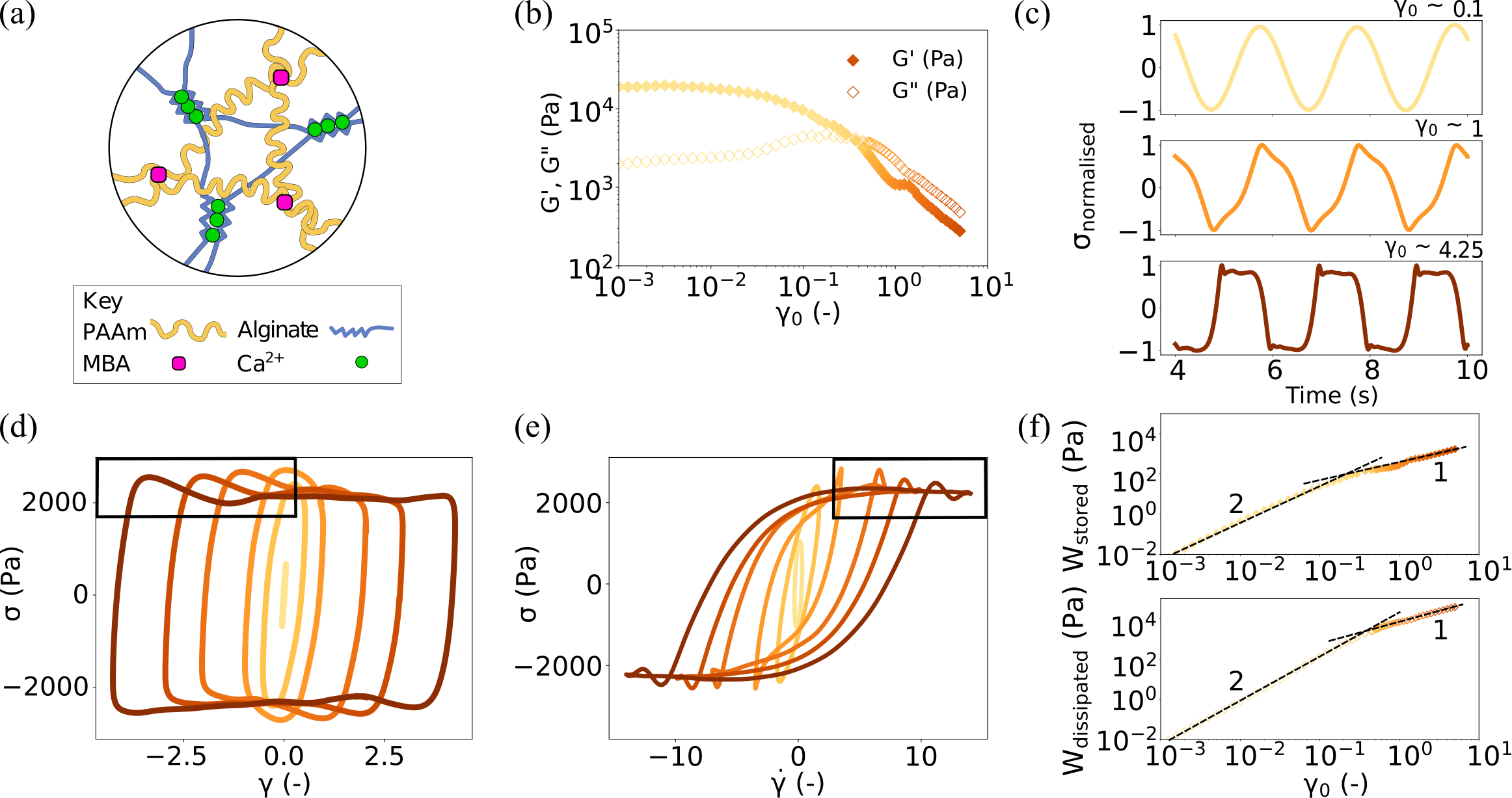}
\caption{\label{figure:IntercycleLB} (a) Schematics of the Alg+/\acrshort{paam}+ hydrogel; (b) Evolution of storage ($G\rq$) and loss ($G\rq$$\rq$) moduli with increasing strain amplitude ($\gamma_{0}$); (c) Representation of the normalised stress response ($\sigma_{normalised}$) to the final cycle of sinusoidal strain with three selected $\gamma_{0}$ highlighting unambiguous changes with increasing amplitude; (d) Elastic \acrshort{lb} projections (stress ($\sigma$) vs strain ($\gamma$)) of the Alg+/\acrshort{paam}+ hydrogel at selected $\gamma_{0}$. The black box highlights "stress overshoot" present in projections of high $\gamma_{0}$; (e) Viscous \acrshort{lb} projections (stress ($\sigma$) vs strain-rate ($\dot{\gamma}$)) of the Alg+/\acrshort{paam}+ hydrogel of selected $\gamma_{0}$. The black box highlights self-intersection present in projections of high $\gamma_{0}$. For both the projections, the $\gamma_{0}$ of oscillation increases from light to darker shades; (f) Evolution of the energy stored ($W_{stored}$) and energy dissipated ($W_{dissipated}$) with increasing $\gamma_{0}$. The dashed lines describe the exponents of the power-law fitted to the adjacent sections on the curves.
}
\end{figure*}

We start by probing the rheological response of a standard Alg+/\acrshort{paam}+ hydrogel (Figure \ref{figure:IntercycleLB}(a)), where both permanent crosslinks and transient crosslinks are present and are expected to contribute to the rheological response. We report the dynamic storage ($G\rq$) and loss ($G\rq$$\rq$) moduli as a function of the applied strain amplitude ($\gamma_{0}$) in Figure \ref{figure:IntercycleLB}(b). At low $\gamma_0$, we observe a characteristic plateau in both dynamic moduli,  signifying the \acrfull{lve} regime of the hydrogel.  Already at a relatively low $\gamma_0$ ($>$ 0.1), a decrease in both $G\rq$ and $G\rq$$\rq$ is observed, indicating nonlinear viscoelasticity and signifying the onset of a breakdown in structure \cite{Hyun2011}. Within this regime, the dynamic moduli lose their strict physical meaning but qualitative inferences can still be made. Notably, $G\rq$$\rq$ exhibits a characteristic weak overshoot as strain accumulates irreversibly \cite{Donley2020} and, at $\gamma_{0}$ = 0.3, a crossover is observed in the dynamic moduli, an approximate signature of the material yielding. Beyond this crossover point, we observe a clear shoulder in the value of $G\rq$ with respect to $\gamma_0$. 

Examining the intra-cycle behavior, in the \acrshort{lve} regime, the stress response ($\sigma$(t)) is initially sinusoidal with a phase shift relative to the strain input ($\gamma$(t)) as seen in Figure \ref{figure:IntercycleLB}(c). In the Fourier transformation of a sinusoidal signal, one harmonic is typically sufficient to accurately represent the rheological response, however, at higher $\gamma_{0}$ ($>$ 0.1) (Figure \ref{figure:IntercycleLB}(c)), the $\sigma(t)$ begins to adopt sinusoidal distortions, which indicates the presence of higher order harmonics \cite{Hyun2011}. We next assess the emergence, type, and extent of intra-cycle non-linear behaviour, by plotting the rheological response as so-called \acrfull{lb} projections as shown in Figures \ref{figure:IntercycleLB}(d) \& \ref{figure:IntercycleLB}(e). Here, the elastic ($\sigma$(t) vs $\gamma$(t)) \acrshort{lb} projections of different $\gamma_{0}$ show that the rheological response adopts an elliptical waveform. Similar elliptical confirmations are seen in the viscous ($\sigma$(t) vs $\dot{\gamma}$(t)) \acrshort{lb} projections (Figure \ref{figure:IntercycleLB}(e)). 

As the sample falls into the non-\acrshort{lve} regime and $\sigma$(t) is distorted, the projections slowly evolve into a curvilinear parallelogram at high strain amplitudes.  To illustrate this example, consider the highest amplitude cycles highlighted in Figure \ref{figure:IntercycleLB}(d) ($\gamma_{0}$ $>$ 1.2).  Here, as the strain increases, the intracycle stress is observed to decrease slightly and then subsequently increase to maximum. This phenomenon is typically referred to as ``stress overshoot'' and indicates structural rearrangement \cite{Benzi2021stressovershoot} of the Alg+/\acrshort{paam}+ hydrogel. This strongly suggests some form of yielding, leading to near-plastic flow. Further supporting evidence of this yielding can be seen in Figure \ref{figure:IntercycleLB}(e) where, within the same range of $\gamma_0$, the viscous \acrshort{lb} projections begin to adopt a sigmoidal shape and show self-intersection in high strain rate cycles.  This self-insertion has been previously linked \cite{Ewoldt2009} to structural rearrangement and, as we observe here, also coincides closely with the stress overshoot. 

To examine the origin of this yielding behaviour, we consider the time-averaged energy stored and the energy dissipation rate, within the Alg+/\acrshort{paam}+ hydrogel over a single LAOS cycle \cite{tschoegl2012phenomenological}
 \begin{equation}
     W_{stored}(\omega)=\frac{1}{4}G'(\omega)\gamma_0^2
 \end{equation}
 \begin{equation}
    W_{dissipated}(\omega) = {\pi}G''(\omega)\gamma_0^2
 \end{equation}
To examine how the total energy is partitioned as the material yields, we consider the increase in $W_{stored}$ and $W_{dissipated}$ as $\gamma_{0}$ increases (Figure \ref{figure:IntercycleLB}(e)). Within the linear regime, both  $W_{stored}$ and $W_{dissipated}$ increase logarithmically with a power law of 2 as expected \cite{tschoegl2012phenomenological}. At higher $\gamma_0$ corresponding to the stress overshoot, the power law relationship of both variables levels out briefly ($\sim$ 0.6, as shown) before increasing again to a value of 1. The clear transition between these two regimes is further evidence of yielding as it indicates a lower increase in energy storage after each oscillatory cycle. This transition in energy stored agrees well with the emergence of a stress overshoot observed in Figure \ref{figure:IntercycleLB}(d) and the self-insertion of \acrshort{lb} viscous projections in Figure \ref{figure:IntercycleLB}(e). 

A number of inferences can be drawn from this transition.  Firstly, the appearance of the stress overshoot in subsequent LAOS cycles indicates that the yielding must be at least partially reversible \cite{Ewoldt2009}. Secondly, the presence of a shoulder at the yielding transition in Figure \ref{figure:IntercycleLB}(b) has previously been observed in a number of studies of transiently crosslinked hydrogels \cite{polym15061558,Wang2023}, although we note that this was not remarked upon by the authors. Finally, the value of $W_{stored}$  continues to increase with increasing $\gamma_0$, after yielding, which indicates that the Alg+/\acrshort{paam}+ hydrogel retains some structural integrity after yielding.  Taken together, this indicates that the yielding at $\gamma_{0}$ $>$ 1.2 is reversible and partial, suggesting a cyclical unbinding and rebinding of the transient Ca$^{2+}$ crosslinks, while the permanent chemical crosslinks between \acrshort{paam} likely remain largely intact. 
\begin{figure*}
\includegraphics[width=\textwidth]{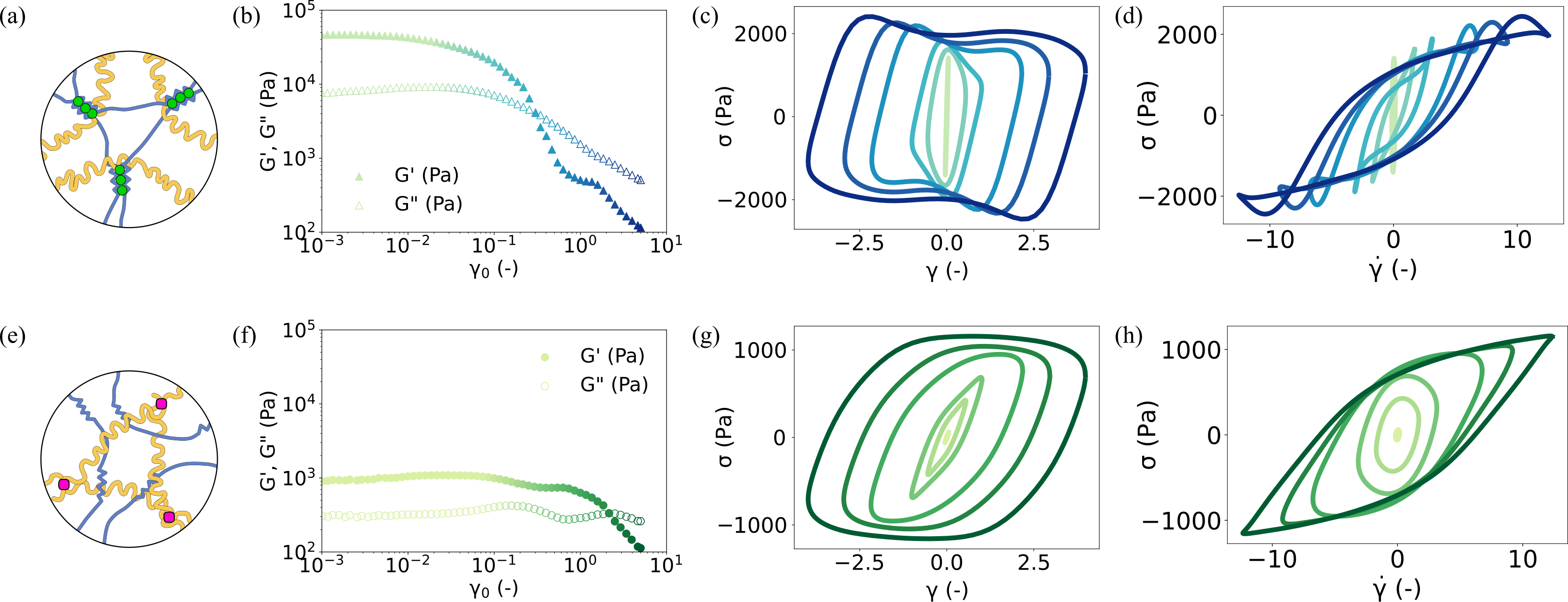}
\caption{\label{figure:IntercycleLBSN} Schematics of structures and behavior under LAOS of the hydrogel designs containing only one network. For Alg+/\acrshort{paam}- hydrogel, (a) illustrates the schematic of the hydrogel with \acrshort{paam} chains uncrosslinked; (b) shows the evolution of $G\rq$ and $G\rq{\rq}$ with increasing $\gamma_{0}$; (c) \& (d) presents the elastic and viscous \acrshort{lb} projections respectively. For  Alg-/\acrshort{paam}+ hydrogel, (e) illustrates the schematic of the hydrogel with the alginate chains uncrosslinked; (f) shows the evolution of $G\rq$ and G$\rq{\rq}$ with increasing $\gamma_{0}$; (g) \& (h) presents the elastic and viscous \acrshort{lb} projections respectively. For all the \acrshort{lb} projections, the $\gamma_{0}$ increases from light to darker shades. 
}
\end{figure*}
\section{Transient Crosslinks Govern the Rheological Response of the Double-Network Hydrogel}

To further examine the role of transient crosslinks in the rheological response of the Alg+/\acrshort{paam}+ hydrogel, we now consider the rheological response where one of two intra-polymer crosslinks is inhibited. To do this, we synthesize composite hydrogels containing both polymers in which the permanent crosslinking is not present (Alg+/\acrshort{paam}- hydrogel) or where the transient crosslinking is not present (Alg-/\acrshort{paam}+ hydrogel) as shown in Figures \ref{figure:IntercycleLBSN}(a) \& \ref{figure:IntercycleLBSN}(e). Performing the identical rheological measurements on both hydrogels,  we observe similar rheological responses between the \acrshort{lve} regime of the Alg+/\acrshort{paam}- hydrogel (Figure \ref{figure:IntercycleLBSN}(b)) and the Alg+/\acrshort{paam}+ hydrogel (Figure \ref{figure:IntercycleLB}(b)), in terms of the $G\rq$ ($\sim10^4$ Pa), $G\rq{\rq}$ ($\sim10^3$ Pa) and the extent of the \acrshort{lve} regime ($\gamma_0 < 0.1$).  Meanwhile, the \acrshort{lve} regime $G\rq$ and $G\rq{\rq}$ of the Alg-/\acrshort{paam}+  hydrogel are approximately an order of magnitude lower, with a \acrshort{lve} regime that extends significantly further ($\gamma_0< 1$) as seen in Figure \ref{figure:IntercycleLBSN}(f). This strongly suggests that the Alg+/\acrshort{paam}+ hydrogel's rheological properties in \acrshort{lve} regime are predominantly governed by the alginate network rather than the PAAm network.
\begin{figure*}
\includegraphics[width=\textwidth]{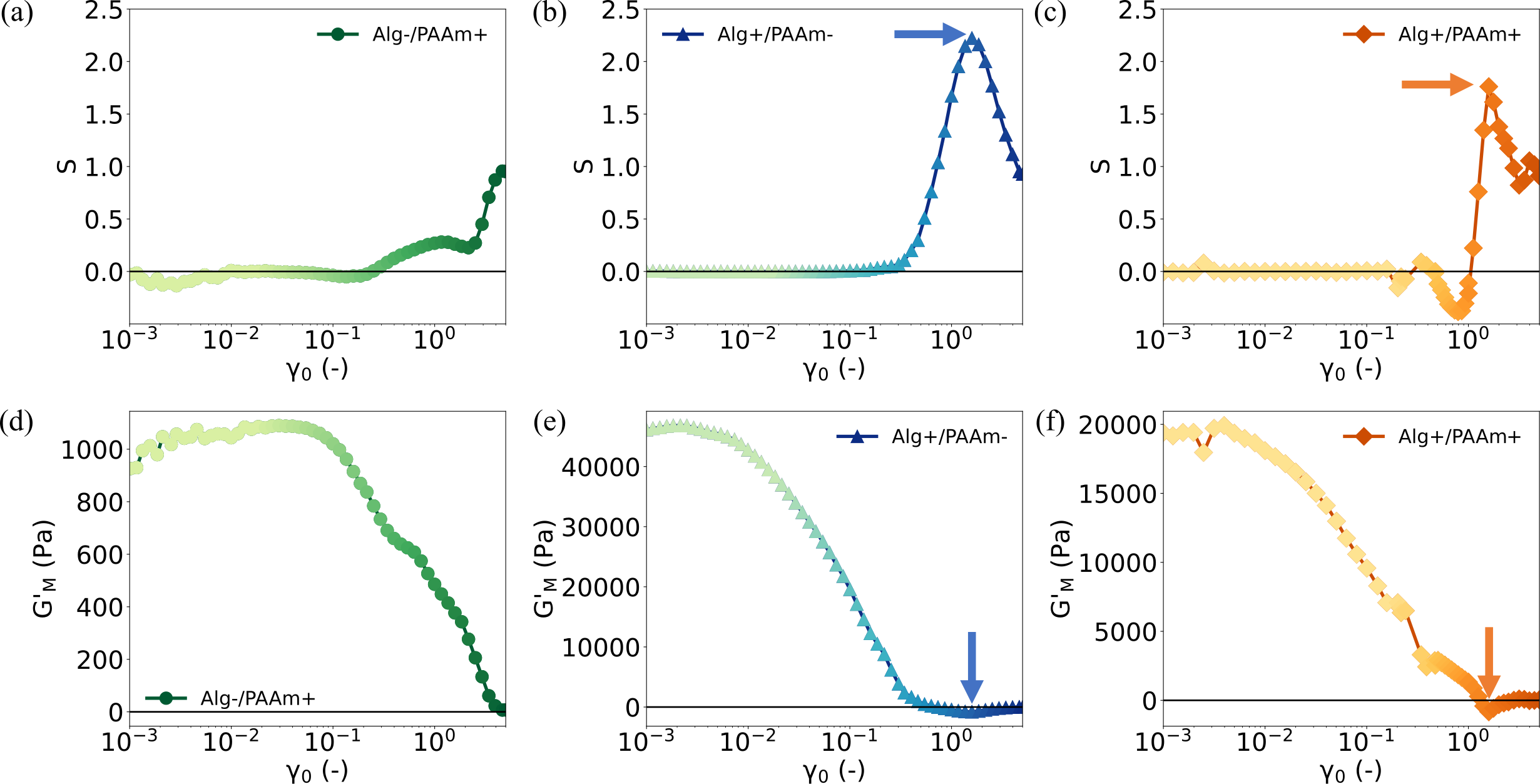}
\caption{\label{figure:ParametersLAOS} Analysis of intra-cycle behaviour of all the hydrogel designs. For Alg-/\acrshort{paam}+, Alg+/\acrshort{paam}- and the Alg+/\acrshort{paam}+ hydrogels, (a), (b), \& (c) show the evolution of strain-stiffening parameter ($S$) while (d), (e), \& (f) show the evolution of zero-strain tangent modulus ($G\rq$$_M$), respectively. The arrows in (b) \& (c) indicate a peak in the $S$ which corresponds to $G\rq$$_M$ taking the most negative values in (e) \& (f). 
}
\end{figure*}
The similarity between the \acrshort{lve} regime of the Alg+/\acrshort{paam}- hydrogel also extends to the nonlinear regime as shown in Figure \ref{figure:IntercycleLBSN}(c). For the Alg+/\acrshort{paam}- hydrogel, the same distorted elliptical elastic \acrshort{lb} projections are observed as seen in Figure \ref{figure:IntercycleLBSN}(c). Notably, at high $\gamma_{0}$, the Alg+/\acrshort{paam}- hydrogel exhibits stress overshoot as seen in the elastic \acrshort{lb} projections, in a manner highly reminiscent of the Alg+/\acrshort{paam}+ hydrogel.  Similarly, the viscous \acrshort{lb} projections of the Alg+/\acrshort{paam} hydrogel show the same characteristic sigmoidal shape as the Alg+/\acrshort{paam}+ hydrogels and exhibit the same self-intersection phenomenon at high strain rates (Figure \ref{figure:IntercycleLBSN}(d)). By contrast, the Alg-/\acrshort{paam}+ hydrogel manifests an entirely different rheological response. We observe an unambiguous presence of plastic flow in the nonlinear response for $\gamma_{0}$ $>$ 1.8, which indicates that yielding has taken place. Notably, stress overshoot is entirely absent from the elastic \acrshort{lb}  projections  (Figure \ref{figure:IntercycleLBSN}(g)), and no self-insertion is observed in the viscous \acrshort{lb} projections (Figure \ref{figure:IntercycleLBSN}(h)). We therefore conclude that these phenomena directly coincide with the presence of transient crosslinks  (Figures \ref{figure:IntercycleLB}(c) \& \ref{figure:IntercycleLB}(d)) and entirely absent when physical crosslinks are inhibited (Figures \ref{figure:IntercycleLBSN}(g) \& \ref{figure:IntercycleLBSN}(h)).  This provides further validation that the nonlinear response of the Alg+/\acrshort{paam}+ hydrogel is a direct consequence of the transient crosslinking of the alginate network.   


\section{The Stiffening Factor Provides a Reliable Measure of Yielding}

Having established the presence of yielding and linked to the alginate crosslink, we now seek to now provide a quantitative perspective on this yielding, by analysing each oscillatory cycle using the commonly used method of decomposing $\sigma$(t) into elastic ($\sigma\rq$) and viscous ($\sigma\rq$$\rq$) components as a series of Chebyshev polynomials of the first kind, $T_n$ \cite{ewoldt2008new,kamkar2022large}. 
\begin{equation}
     \sigma\rq(x : \omega,\gamma_0) = \gamma_0 \sum e_n(\omega,\gamma_0) T_n(x)
 \end{equation}
\begin{equation}
\sigma\rq{\rq}(y : \omega,\gamma_0) = \dot{\gamma_0} \sum v_n(\omega,\gamma_0) T_n(y) 
\end{equation}
where, $x$ = $\gamma$/$\gamma_0$ and $y$ = $\dot{\gamma}$/$\gamma_0$ depict the normalised strain and strain-rate. The coefficients $e$ (unit: Pa) and $v$ (unit: Pa$\cdot$s) represent elastic and viscous contributions respectively. With increasing order, the magnitude of each Chebyshev coefficient decreases monotonically. A physical interpretation of the nonlinearity can be reached by observing the sign of $e_3$ and $v_3$ since they determine the concavity of the $\sigma\rq$ and $\sigma\rq$$\rq$.  Changes in these parameters, within and between cycles, reflect changes in the rheological response \cite{Wang2022,kamkar2022large}.    Within cycles, this change can be conveniently quantified through the  stiffening factor: 
\begin{equation}
    S =  (G\rq_L-G\rq_M)/G\rq_L 
\end{equation}
where $G\rq$$_L$ is a large amplitude secant slope and $G\rq$$_M$ is a zero amplitude tangent slope as highlighted in Figure S3 in Supplemental Material \cite{supp}. Put simply, $G\rq$$_L$ probes the rheological response at the maximum strain in a cycle, while  $G\rq$$_M$ characterises the response at zero strain in a cycle. A value of $S = 0$, indicates $G\rq$$_M$ =$G\rq$$_L$ and is therefore indicative of a \acrshort{lve} response; $S$ $>$ 0 is taken to indicate intracycle strain stiffening, and  $S$ $<$ 0 is taken to indicate intracycle strain-softening. 

Figures \ref{figure:ParametersLAOS}(a)-(c) show the stiffening factor $S$ with respect to $\gamma_0$ for each Alg+/\acrshort{paam}+ hydrogel design. In all three hydrogel designs, the value of $S$ is close to zero at low values of $\gamma_{0}$ corresponding to the \acrshort{lve} regime, as expected.  At higher $\gamma_0$, all hydrogel designs exhibit a slight temporary decrease in $S$ followed by a sustained increase,  signifying intra-cycle strain stiffening. The resulting increase in $S$ is highest for samples in which alginate is crosslinked (Figures \ref{figure:ParametersLAOS}(b) \& \ref{figure:ParametersLAOS}(c)) and in both cases reaches a peak at high $\gamma_0$, before decreasing. Meanwhile, when alginate is not crosslinked, stiffening extends to the end of the experimentally accessible range in $\gamma_0$ (Figure \ref{figure:ParametersLAOS}(a)).  It is likely that this intra-cycle strain-stiffening arises from the semiflexible nature of the alginate polymer, which results in entropy-driven stiffening as the polymer chains align in the direction of shear \cite{storm2005nonlinear}. Notably, similar behaviour has been seen in hydrogels constructed from pectin \cite{John2019}, which has a similar molecular structure to alginate. The subsequent decrease in $S$ in the presence of alginate crosslinking may be rationalised by the yielding of the alginate network and dissociation of crosslinks \cite{John2019,polym15061558}.  The transition between stiffening and yielding is therefore characterised by a peak in $S$  and therefore provides a convenient way to clearly identify the yielding of the alginate network.  

Another rheological metric that makes the failure evident is the value of $G\rq$$_M$, as shown for each of the three hydrogel designs in Figures  \ref{figure:ParametersLAOS}(d)-(f).  Here, it can be clearly seen that $G\rq$$_M$ becomes negative when $S$ peaks, and furthermore, that maximum in $S$ corresponds to a minimum in $G\rq$$_M$.  From an examination of the elastic \acrshort{lb} projections in Figures \ref{figure:IntercycleLB}(d) and \ref{figure:IntercycleLBSN}(c), it is clear that a negative value of $G\rq$$_M$  is synonymous with the onset of the negative local slope and stress overshoot.  This provides further evidence that the peak in $S$ correlates closely with the yielding of the alginate network. Interestingly, in the case of Alg-/\acrshort{paam}+ hydrogel (Figure \ref{figure:ParametersLAOS}(a)), we observe a positive value of $G\rq$$_M$ across the entire measurement range, which is consistent with the absence of alginate crosslinking.  Despite this, some strain stiffening can be seen in Figure \ref{figure:ParametersLAOS}(d), albeit over a longer range of $\gamma_0$ and to a lesser extent. 

It should be noted that the conventional amplitude sweep data in Figures \ref{figure:IntercycleLB}(b), \ref{figure:IntercycleLBSN}(b) and \ref{figure:IntercycleLBSN}(e) would indicate that all hydrogel designs exhibit softening, whereas the positive values of  $S$ in Figure \ref{figure:ParametersLAOS}(a)-(c) gives the paradoxical view of the hydrogels exhibiting strain-stiffening.  To resolve this, the elastic stress for the system needs to be examined \cite{MermetGuyennet2014}. The elastic stress is observed to decrease in successive oscillatory cycles, (Figure S4 in Supplemental Material \cite{supp}) which confirms that the hydrogels undergo gradual softening.  We therefore surmise that the parameters $S$, $G\rq$$_M$, and $G\rq$$_L$, reflect only the rheological response within a single oscillatory cycle. This makes them beneficial only for characterising the \acrshort{lb} projections and great care must be taken when using them to make physical interpretations \cite{Ewoldt2013}. Therefore we may say that all Alg+/\acrshort{paam}+ hydrogel designs undergo strain stiffening within individual oscillatory cycles but undergo softening after repeated deformation.  One final point to support this is that, at large $\gamma_{0}$, $G\rq$$_M$ becomes vanishingly close to zero due to the presence of plastic flow at zero-strain as compared to $G\rq$$_L$ which makes the value of $S$ close to 1 inherently. 

\section{PAAm Network Supports the Structure of the Double-Network Hydrogel}

\begin{figure*}
\includegraphics[width=\textwidth]{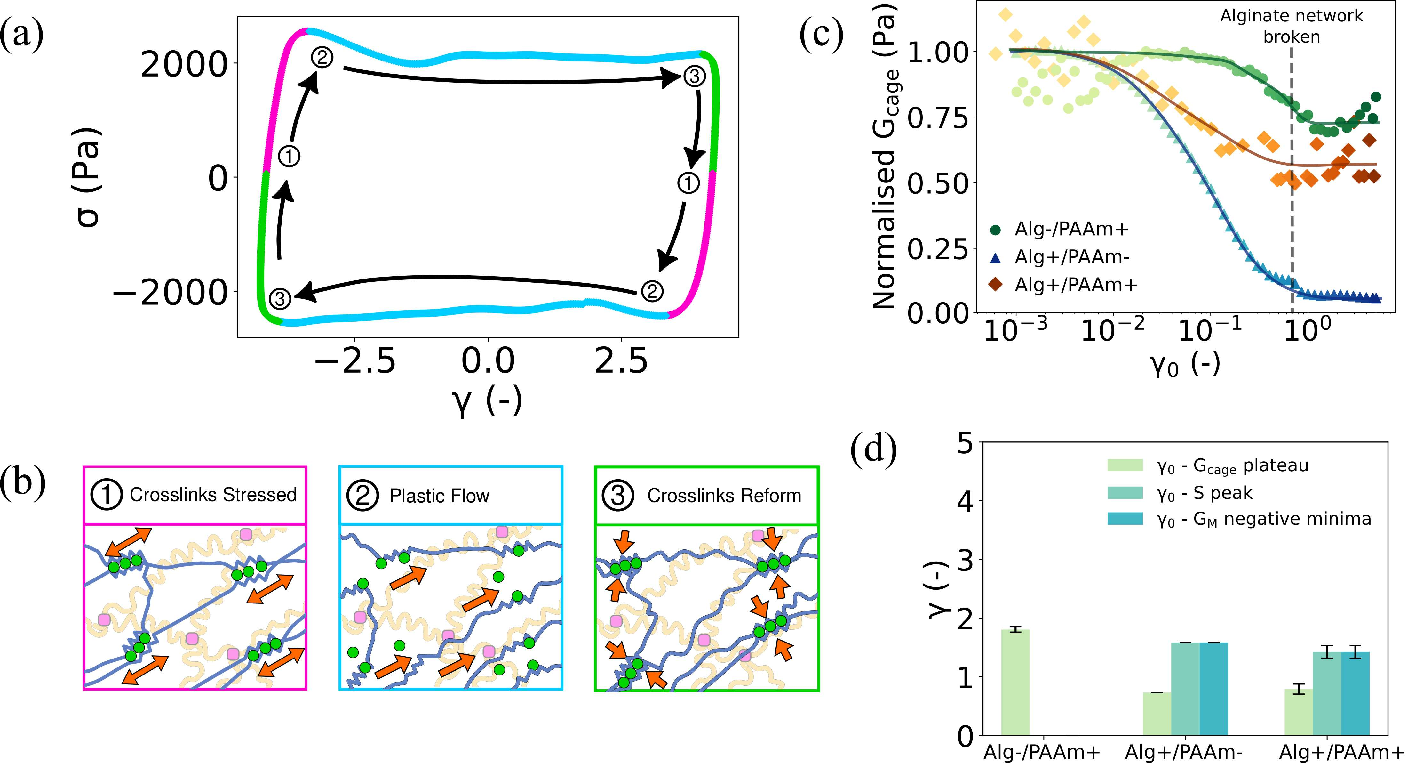}
\caption{\label{figure:CageModulus} (a) illustrates the sequence of physical events that Alg+/\acrshort{paam}+ hydrogel undergoes in a cycle of large amplitude deformation ($\gamma_{0}$ $\sim$ 4.25); (b) presents the schematics of microstructural events in the Alg+/\acrshort{paam}+ hydrogel corresponding to different highlighted parts on the elastic \acrshort{lb} projection presented in (a); (c) shows the evolution of normalised cage modulus (\acrshort{gcage}) for all the hydrogels. The cage modulus has been normalised by the average cage modulus observed in the \acrshort{lve} regime of the hydrogels and guide-to-eye solid lines are drawn to highlight different features in the evolution. The dashed line highlights the onset of the second plateau in \acrshort{gcage} at high $\gamma_{0}$; (d) presents $\gamma_{0}$ values for the point at \acrshort{gcage} starts to plateau at high deformations, the point at which peak in $S$ is observed, and the point at negative minima in $G\rq$$_M$ is observed for all hydrogel designs.
}

\end{figure*}

Having explored the role of the alginate network in the hydrogel's rheological response, we next examine the contribution of the \acrshort{paam} network. We have shown that this network remains largely intact across the experimental range of strain amplitudes. To probe the structural changes in the hydrogel, as the alginate network yields, we consider the local behaviour of the network at each stage of the oscillatory shear cycle, as shown in Figure \ref{figure:CageModulus}.  The schematic shows a representative oscillatory cycle carried out on  Alg+/\acrshort{paam}+ hydrogel at $\gamma_0 \sim$ 4.25, an amplitude at which the alginate network yields but the \acrshort{paam} network remains intact. 

It is important to note that this oscillatory cycle exhibits 180$^{\circ}$ rotational symmetry around the origin. As with all LAOS waveforms \cite{Ewoldt2009}, this reflects that the underlying structural changes in the material are fully reversible within each cycle. Examining this waveform and beginning at zero shear stress, the hydrogel first exhibits elastic strain as the sample is sheared (highlighted by the pink curve segment; \textcircled{1} $\rightarrow$ \textcircled{2}).  This initial stage reflects the alignment of the alginate polymer chains in the direction of shear as the crosslinks are placed under tension (Figure \ref{figure:CageModulus}(b)). When the strain is increased further, the stress overshoot is reached (\textcircled{2}) at which point the alginate network's ionic crosslinks unbind, leading to a decrease in stress. The breakdown of the alginate network structure induces plastic flow (as highlighted by the blue curve segment; \textcircled{2} $\rightarrow$ \textcircled{3}). Following this, \textcircled{3} denotes the point at which the strain rate is zero and $\gamma = \pm\gamma_{0}$. At this point, the shear stress relaxes (\textcircled{3}) as the direction of strain is reversed and the crosslinks re-bind to cause the alginate network to reform (green curve segment; \textcircled{3} $\rightarrow$ \textcircled{1}) likely due to strong interaction between Ca\textsuperscript{2+} ions and G-units of the alginate chains. 

These microstructural changes can be inferred by computing the gradient of the raw $\sigma$(t)-$\gamma$(t) curve at zero stress.  This variable, commonly referred to as the \textit{cage modulus}, $G_{cage}$, reflects the rheological response of the hydrogel at mechanical equilibrium within each oscillatory cycle since the elastic and viscous stresses at this point are either zero or momentarily have equal magnitudes with opposite signs \cite{Rogers2011}. The evolution of $G_{cage}$, normalised by its initial value,  with increasing $\gamma_{0}$ is presented in Figure \ref{figure:CageModulus}(c) for all three hydrogel designs. In the \acrshort{lve} regime, $G_{cage}$ for all the hydrogels does not vary with $\gamma_0$ and aligns closely with measured values of $G\rq$ (Figure S6 in Supplemental Material \cite{supp}). This indicates that no microstructural changes are observed in \acrshort{lve} regime, as expected. As $\gamma_{0}$ increases, $G_{cage}$ for all the hydrogels begins to decrease, indicating microstructural breakdown. 

The nature of this breakdown varies substantially between hydrogel designs. When alginate is crosslinked (Alg+/\acrshort{paam}- and Alg+/\acrshort{paam}+ hydrogels), the observed decreases in $G_{cage}$ occur at a similar strain amplitude ($\gamma_{0}$ $\sim$ 0.01) and at similar initial rates.  All hydrogel designs also exhibit a plateau in the normalised $G_{cage}$,  at high $\gamma_{0}$. This indicates that there is an upper limit the elastic energy that is stored by the network \cite{Rogers2023} and implies that any further strain imposed on the hydrogel results only in energy dissipation. Where only alginate is crosslinked (Alg+/\acrshort{paam}- hydrogel), the value of $G_{cage}$ reduces by over 90$\%$ of its initial value suggesting a catastrophic microstructural breakdown following yielding.  

To understand the role of the \acrshort{paam} in the yielding process we compare the evolution of $G_{cage}$ between the three hydrogel designs. Comparing Alg+/\acrshort{paam}- and Alg+/\acrshort{paam}+ hydrogels, we note that the presence of \acrshort{paam} crosslinking substantially lessens the structural breakdown since the value of $G_{cage}$ reduces to only $\sim$50$\%$ of its initial value. From this, we infer that, for the Alg+/\acrshort{paam}+ hydrogel, once the alginate network breaks, the \acrshort{paam} network maintains the integrity of the hydrogel and inhibits the macroscopic failure.   

Figure \ref{figure:CageModulus}(d) summarises the key features of network failures identified by both modes of analysis presented in the study. The onset of this plateau occurs at a lower $\gamma_0$ to the point at yielding occurs as highlighted by the peak in $S$ and the negative value of $G\rq_M$ (Figure \ref{figure:ParametersLAOS}).  Thus, we surmise that the structural breakdown reflected by $G_{cage}$ can be considered the initial stage of the yielding transition. We note that the point of onset of $G_{cage}$ plateau is somewhat subjective while the peak in $S$ can be evaluated precisely. Due to its lack of subjectivity and direct correlation with the yield point, the point of peak in $S$ determines the point of yielding of the alginate network in Alg+/\acrshort{paam}- and Alg+/\acrshort{paam}+ hydrogels precisely.

The most common general hypothesis for the breakdown of double-network hydrogels under strain posits that the more brittle network (in this case, alginate) fragments into clusters and that these clusters prevent further structural breakdown of the remaining network (in this case, \acrshort{paam}) by acting as secondary crosslinks \cite{Gong2010}. If this were the case here, we would expect that the absence of a crosslinked alginate network would lead to a more dramatic structural breakdown. Instead, we observe the opposite.  Where only \acrshort{paam} is crosslinked (Alg-/\acrshort{paam}+), the reduction in normalised $G_{cage}$ is less severe. Our findings are therefore not consistent with the cluster theory, although it should be emphasised that further structural breakdown might occur at higher $\gamma_0$, beyond the experimentally accessible range of the rheometer. This would be consistent with the ductile nature of the \acrshort{paam} network \cite{xu2018decoupling,Gong2010}. In either case, our data  suggests that the mechanical role of the alginate network is to dissipate energy through continuous yielding and re-forming, while the role of the \acrshort{paam} network is to retain the Alg+/\acrshort{paam}+ hydrogel's structural integrity and prevent macroscopic failure. 

\section{Hydrogels containing Transient Crosslinking Exhibit Two-Step Yielding}
 \begin{figure}
\includegraphics[width=0.86\columnwidth]{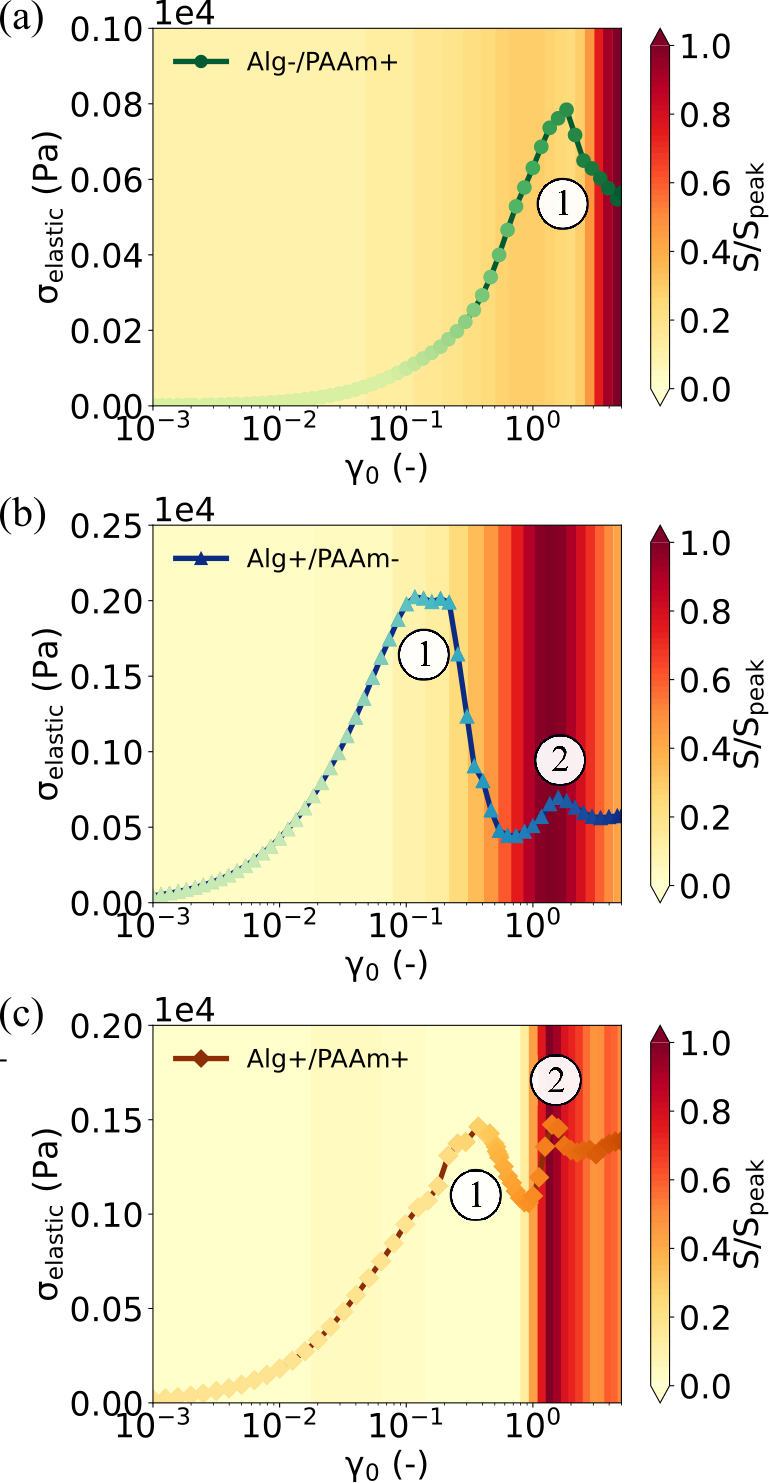}
\caption{\label{figure:StressContribution} Determination of yield point. (a), (b), and (c) demonstrate the elastic contribution to the total stress ($\sigma_{elastic}$)  experienced by the Alg-/\acrshort{paam}+, Alg+/\acrshort{paam}-, and Alg+/\acrshort{paam}+ hydrogels respectively. The peaks in $\sigma_{elastic}$ are numbered as (1) \& (2) in the order of occurrence. 
}

\end{figure}

It should be noted that our use of peak stiffening factor $S$ as a signature of yielding is not a standard metric to quantify yielding. Indeed there are many approaches to infer a single yield point in the \acrshort{laos} curves \cite{Dinkgreve2016}, including the onset of non-linearity and the crossover between $G\rq$ and $G\rq$$\rq$. While these definitions are instructive for identifying trends between analyses, they are typically not consistent with one another \cite{Walls2003} and may therefore not offer a true representation of the yield point \cite{Donley2019}. A more reliable approach is to calculate the contribution of elastic stress to the total stress ($\sigma_{elastic}$ $\sim$ $G\rq$$\gamma_{0}$), from which the yield point may be interpreted as the point at which elastic stress decays \cite{Yang1986, Pan2003, Walls2003}. The key advantages of this approach are that it can offer a precise estimation of the yield point, it is less subjective and it is consistent with conventional steady shear measurements. 

To support our claim that the $S$ peak is a reliable measure of yielding, we independently quantify the yield point by this method, plotting the $\sigma_{elastic}$ with respect to $\gamma_0$ for all three hydrogel designs in Figure \ref{figure:StressContribution}. The colour bars in the figures represent the values of $S$, extracted from Figure \ref{figure:ParametersLAOS}, normalised by their maximum values.  As expected, for hydrogel designs in which alginate is crosslinked (Figures \ref{figure:StressContribution}(b) \& \ref{figure:StressContribution}(c)) a clear local maximum (labelled (2)) is observed in $\sigma_{elastic}$ that corresponds directly to the measured peak in $S$. 

Remarkably, we also observe a peak in $\sigma_{elastic}$ for the Alg-/\acrshort{paam}+ hydrogel, in which alginate crosslinks are not present and an additional peak for hydrogels containing transient crosslinks. In both cases, this peak appears (labelled (1) in Figures \ref{figure:StressContribution} (a)-(c)) at a value of $\gamma_0$ than is approximately one order of magnitude below the yield point (2). Examining the relative contribution of the viscous stress (Figures S5(a)-(c) in the Supplemental Material \cite{supp}), we note that the viscous contributions begin to increase immediately following this initial peak.  This therefore indicates that all three hydrogel designs undergo an initial and separate yielding event that is not clearly manifested in the strain amplitude sweep data (Figures \ref{figure:IntercycleLB}(b), \ref{figure:IntercycleLBSN}(b) \& \ref{figure:IntercycleLBSN}(f)) nor in the \acrshort{lb} projections (Figures \ref{figure:IntercycleLB}(d), \ref{figure:IntercycleLB}(e), \ref{figure:IntercycleLBSN}(c), \ref{figure:IntercycleLBSN}(d), \ref{figure:IntercycleLBSN}(g) \& \ref{figure:IntercycleLBSN}(h)). 

The presence of two-step yielding events has been observed for numerous systems \cite{Ahuja2020} and arises when $\sigma_{elastic}$ relaxes twice \cite{Ahuja2020} in a similar manner to that seen in Figures \ref{figure:StressContribution}(b) \& \ref{figure:StressContribution}(c). In our system, the second peak (2) coincides with the $S$ peaks for hydrogel designs in which alginate is crosslinked,  making it evident that the failure in the alginate network triggers yielding. The mechanistic origins of the first peak (1) are less obvious, however. In prior studies of two-step yielding, the emergence of a second peak is generally thought to arise from the presence of a second distinct yielding mechanism \cite{Ahuja2020}. Examples include cluster breaking in colloidal gels \cite{koumakis2011two}, cage-breaking in glassy systems \cite{pham2008yielding}, cage-breaking over different length scales in bi-disperse suspensions \cite{sentjabrskaja2013yielding}, and aggregate remodelling in fibrous networks \cite{xu2018decoupling}. Notably, the emergence of two-step yielding is also often correlated with the presence of an additional component \cite{matsuda2016yielding} or a new inter-molecular interaction \cite{xu2018decoupling,potanin2019rheology}. Considering the components in our hydrogel designs, our interpretation already accounts for the yielding of the alginate network and, furthermore,  the observed increase in the energy stored across the experimental range (Figure \ref{figure:IntercycleLB}(f)) indicates that the PAAm network does not yield. Considering possible chemical interactions, the alginate crosslinking and polyacrylamide crosslinking can be discounted for the same reasons.  However, there exists a final possible interaction, which is the intermolecular interactions \textit{between} the two polymer networks. Prior spectroscopic studies of similar hydrogel designs have compared the adsorption bands of composite and single-component gels and shown that the combination of PAAm and alginate in hydrogels led to the formation of hydrogen bonds, likely between the -NH\textsubscript{2} groups of PAAm and the -COO\textsuperscript{-} groups of alginate \cite{Sun2012}.  If such hydrogen bonds were present, we would expect them to be weaker than the ionic bonds that crosslink the alginate chains, which is consistent with the presence of the first peak at a lower strain (1) than the second peak (2). It would explain the consistent presence of the additional peak across all three hydrogel designs, across which hydrogen bonding \textit{between} the polymer networks is the only common factor.  
\begin{figure*}[!htb]
\includegraphics[width=\textwidth]{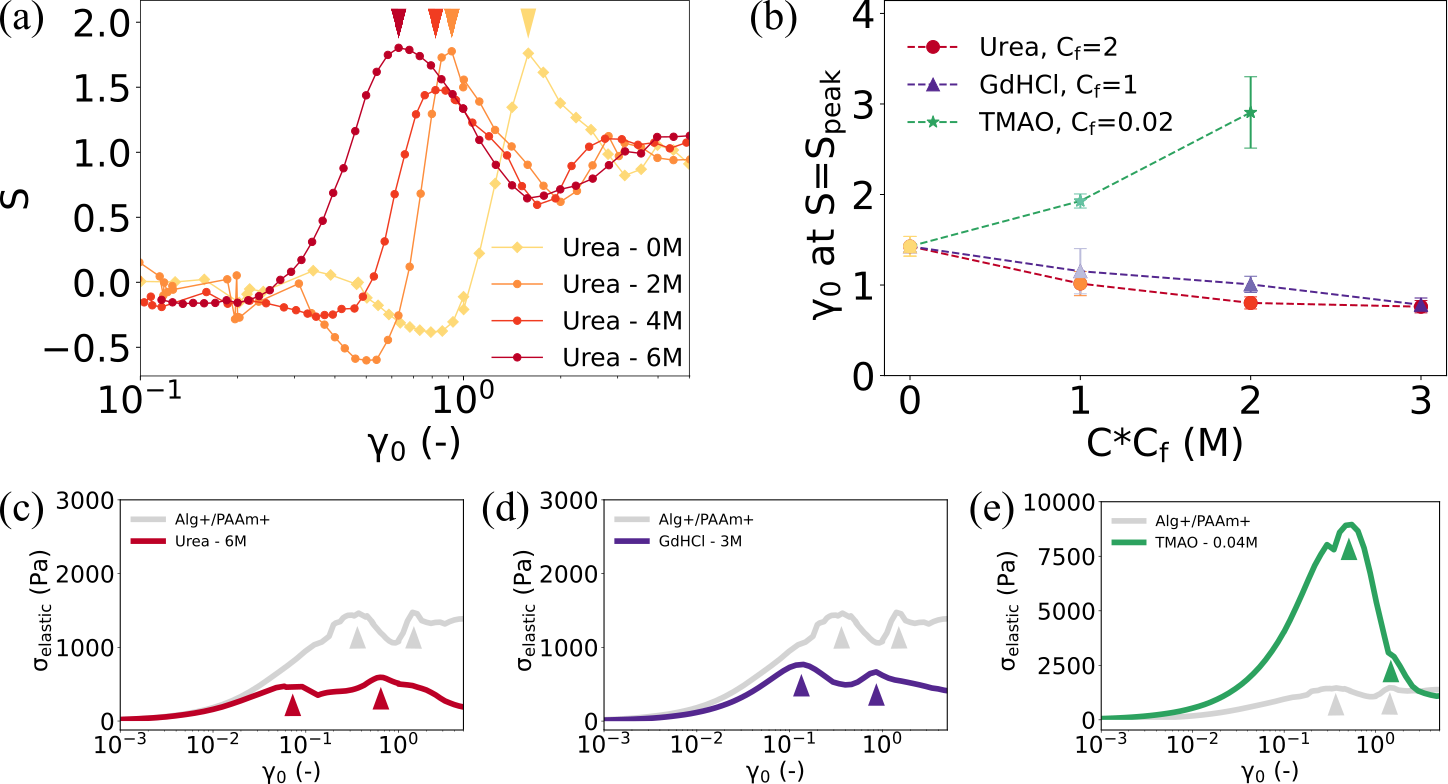}
\caption{\label{figure:Internetwork} Demonstration of effects of changing the strength of inter-network crosslinks. (a) shows the shift of peak in $S$ to lower values of $\gamma_{0}$ with increasing urea concentration in the solution in which Alg+/\acrshort{paam}+ hydrogel is soaked; (b) shows the evolution of peak shift with increasing urea, \acrshort{gdhcl}, and \acrshort{tmao} concentration. A scaling parameter ($C_{f}$) is introduced. The actual concentration can be calculated by multiplying the x-value of the data point with $C_{f}$; (c), (d) \& (e) present the evolution of $\sigma_{elastic}$ with increasing $\gamma_{0}$, for Alg+/\acrshort{paam}+ hydrogel soaked in 6 M urea, 3 M \acrshort{gdhcl}, and 0.04 M \acrshort{tmao} solutions, respectively. The grey curve represents the evolution of $\sigma_{elastic}$ with increasing $\gamma_{0}$ of Alg+/\acrshort{paam}+ hydrogel untreated with any of the agents.
}
\end{figure*}

\section{Hydrogen Bonding Provides the Mechanism for Two-Step Yielding}

Having established hydrogen bonding between the alginate and \acrshort{paam} polymer networks as a plausible mechanism for the two-step yielding shown in Figure \ref{figure:StressContribution}, we now test this hypothesis.  Investigating the role of hydrogen bonding requires a method to adjust the bond strength. We, therefore, introduce two chaotropic reagents, urea and \acrshort{gdhcl} to our hydrogel designs. Both reagents are expected to disrupt the hydrogen bonds between the two polymer networks of the Alg+/\acrshort{paam}+ hydrogel as they preferentially interact with hydrogen binding sites \cite{UreaHbond, Raskar2019}. As a further control, we also introduce the kosmotropic reagent \acrshort{tmao}, which bonds with water molecules and decreases their diffusivity.  This ultimately increases the lifetime of the native hydrogen bonds, thereby providing reinforcement \cite{arakawa1985stabilization}.

As before, we quantify the yielding point by identifying a peak in the $S$ parameter. Figure \ref{figure:Internetwork}(a) shows the evolution of $S$ with the increase in $\gamma_{0}$ at different urea concentrations for a hydrogel design in which alginate and \acrshort{paam} are both crosslinked. Remarkably, the peak in $S$ is observed to shift to a lower $\gamma_{0}$ value as the urea concentration increases, indicating that the hydrogel undergoes yielding more readily when hydrogen bonds are inhibited. Figure \ref{figure:Internetwork}(b) depicts the $S$ peak shift effect for \acrshort{gdhcl}, and \acrshort{tmao} alongside urea. Importantly, the addition of \acrshort{gdhcl} also results in yielding at a lower strain, which confirms that this phenomenon is a direct result of reduced hydrogen bonding, rather than any other chemical reaction associated individually with urea or \acrshort{gdhcl}.  Furthermore, the $S$ peak is approximately twice as sensitive to  \acrshort{gdhcl} concentration compared to urea concentration. This is seen by the reduction in the $S$ peak from 1.42 to 1.16 in the presence of 1 M \acrshort{gdhcl} and from 1.42 to 1.00 in the presence of 2 M urea.  The relative ability of \acrshort{gdhcl} and urea to disrupt hydrogen bonds in monomeric proteins is generally accepted to be 2:1 \cite{Myers1995, Pace2000}, providing further evidence that the observed shift in yielding position is linked to the disruption of hydrogen bonds. 

By contrast, the reinforcement of hydrogen bonds through the introduction of \acrshort{tmao} increases the strain at which $S$ peaks. Here, the $S$ peak shifts from 1.42 when no TMAO is present, to 1.92 when 0.02 M \acrshort{tmao} is added and to 2.90 when 0.04 M \acrshort{tmao} is added. As a separate measure of yielding, we also plot the evolution of of the $\sigma_{elastic}$, examining the original two-step yielding (Figure \ref{figure:StressContribution}(c)) in comparison to the equivalent hydrogel design with the inclusion of urea (Figure \ref{figure:Internetwork}(c)), \acrshort{gdhcl} (Figure \ref{figure:Internetwork}(d)) and \acrshort{tmao} (Figure \ref{figure:Internetwork}(e)).  As expected the additional peak in $\sigma_{elastic}$ at low strain is reduced in the presence of urea and \acrshort{gdhcl} and increased in the presence \acrshort{tmao} (Figures \ref{figure:Internetwork}(c)), \ref{figure:Internetwork}(d), \ref{figure:Internetwork}(e), \& S7 in Supplemental Material \cite{supp}).  Interestingly the second peak, which we ascribe to alginate network yielding, is also reduced when hydrogen bonds are inhibited and increased when hydrogen bonds are reinforced.  This indicates that the hydrogen bonds play a central role in the rheological response: their presence allows more energy to be stored elastically before the yielding of the alginate network and their absence limits the ability of the alginate network to store elastic energy. 

Overall, our findings indicate a synergy between the two polymer networks that is crucial to understanding the overall rheological response, since the hydrogen bonds between the two polymer networks directly determine the material's yielding behaviour. Based on these findings it appears that the presence of hydrogen bonds influences the partitioning of elastic energy between the two polymer networks. In the presence of hydrogen bonds, the energy is stored in both polymer networks, and a yielding of the hydrogen bonds is observed before the yielding of the alginate network (Figure \ref{figure:StressContribution}(a)).  When hydrogen bonds are inhibited, the energy is stored predominantly in the alginate network, such that the mechanical load is not shared between the polymer networks but rather is applied to the alginate polymer chains, leading to the alginate network yielding at lower strain (Figure \ref{figure:Internetwork}(a)). Conversely, the reinforcement of the hydrogen bonds enhances the propensity of the two polymer networks to share elastic stress and the hydrogel undergoes yielding at significantly higher strains. 

\section{Conclusion}

We have carried out a systematic synthesis of different hydrogel designs constructed from interpenetrating polymer networks of alginate and \acrshort{paam} polymer chains and thereby identified the mechanistic origins of their yielding using \acrshort{laos} rheology. This systematic approach is beneficial particularly as it allows direct comparison of mechanical behaviour with network structural parameters and we believe it may be extended to study a plethora of other polymeric composites. 

The model system examined here undergoes a process of two-step yielding when the strain amplitude is increased, provided that the alginate network is crosslinked. This is among very few polymeric hydrogels that have been shown to exhibit the phenomena \cite{matsuda2016yielding,xu2018decoupling}, however, this may reflect the fact that very few such rheological studies have been carried out on composite hydrogels and, as we argue, probable signatures of two-step yielding may have been overlooked \cite{polym15061558,Wang2023}. We believe the quantification of this two-step yielding, may present a powerful approach to identifying how polymeric composites undergo yielding. In this study, we show that the first step of this yielding is governed by the breakage of hydrogen bonds between the two polymer networks, while the second reflects the unbinding of ionic crosslinks between the alginate chains. Significantly, both yielding steps are shown to be reversible, as evidenced by the presence of repeated stress overshoot signatures and self-insertion in the \acrshort{lb} projections (Figure \ref{figure:IntercycleLB}).  Our data indicates that, provided both polymer networks are crosslinked, the rebinding of these crosslinks occurs within each oscillatory cycle (2 seconds) indicating that the hydrogels are able to rapidly self-repair. This provides a potential mechanism for the previously reported ability of such hydrogels to dissipate energy under strain \cite{chen2014fracture} and recover after large deformations \cite{Sun2012}. We have also shown that the alginate network determines the rheological properties in \acrshort{lve} regime while the largely intact presence of the \acrshort{paam} network (Figure \ref{figure:CageModulus} (c)) indicates its role in maintaining the microstructure of the Alg+/\acrshort{paam}+ hydrogel after the alginate network ruptures. To our knowledge, this is the first direct rheological evidence to support the hypothesis that a softer and ductile network prevents catastrophic failure in a double-network hydrogel by acting as a sacrificial network \cite{Gong2010}.

Our results also highlight a new distinct approach in which the yielding of one network within a composite hydrogel may be identified, through the identification of a peak in the stiffening factor $S$.  This reflects the semiflexible nature of the alginate chains, which exhibit entropic stiffening as the chains are placed under tension when sheared \cite{storm2005nonlinear}. Meanwhile the flexible nature of the \acrshort{paam} chains means they do not exhibit stiffening, providing a convenient approach to identify the yielding as the point at which the mechanical load transfers from the semiflexible alginate network to the flexible \acrshort{paam} network. Remarkably, we show that this yield point is sensitive to the hydrogen bonds that govern the interactions between the two polymer networks. Adding chaotropic reagents that weaken the strength of the hydrogen bonds results in hydrogels that yield at lower $\gamma_{0}$ while cosmotropic reagents are able to delay the yielding of the alginate network. It will be interesting to examine whether similar trends in yielding emerge from other composite hydrogels which include semiflexible networks such as agar \cite{chen2014fracture} or collagen \cite{wang2018preparation}. 

Overall, we have combined a systematic synthesis approach with the quantitative analysis of oscillatory rheology data through \acrshort{laos} rheology.  Through this approach, we have found that it is possible to present a nuanced and detailed analysis of rheological behaviour and underlying microstructural changes in composite hydrogels. From the perspective of polymer composites, we have shown that crosslinks \textit{between} individual polymer networks are as important as the crosslinks \textit{within} each network. Thus, controlling the inter-network crosslinking density or strength can be considered as a design parameter for composites that can more closely mimic the mechanical properties of natural tissues. 

\section{Acknowledgements}

V. K. is supported by an Engineering and Physical Sciences Research Council (EPSRC) PhD studentship through the Soft Matter and Functional Interfaces Centre for Doctoral Training (SOFI CDT) (EP/L015536/1), co-funded by Reckitt, UK.  We appreciate the valuable insights and suggestions offered by Gareth McKinley and Simon Rogers. We thank Adrian Hill from Netzsch for help with the rheology protocol design and optimisation, and Alan Smith for kindly lending a serrated parallel plate. We also thank Randy Ewoldt for sharing the MITlaos software.

\bibliographystyle{apsrev4-1} 

\begin{quotation}

\end{quotation}
\bibliography{prl}
\newpage
 
\renewcommand{\thesection}{S\arabic{section}}  
\renewcommand{\thetable}{S\arabic{table}}  
\renewcommand{\thefigure}{S\arabic{figure}} 
\setcounter{figure}{0}
\onecolumngrid

\newpage

\begin{center}
    \Large{Supplementary Information for}\\
     \vspace{12pt}
    \Large{\textbf{Mechanistic Origins of Yielding in Hybrid Double Network Hydrogels}}\\
    \vspace{12pt}
    \large{Vinay Kopnar, Adam O'Connell, Natasha Shirshova and Anders Aufderhorst-Roberts}
\end{center}

\vspace{12pt}

\setcounter{section}{0}
\section{The custom-designed mould}
\begin{figure*}[!htb]
\includegraphics[width=0.9\textwidth]{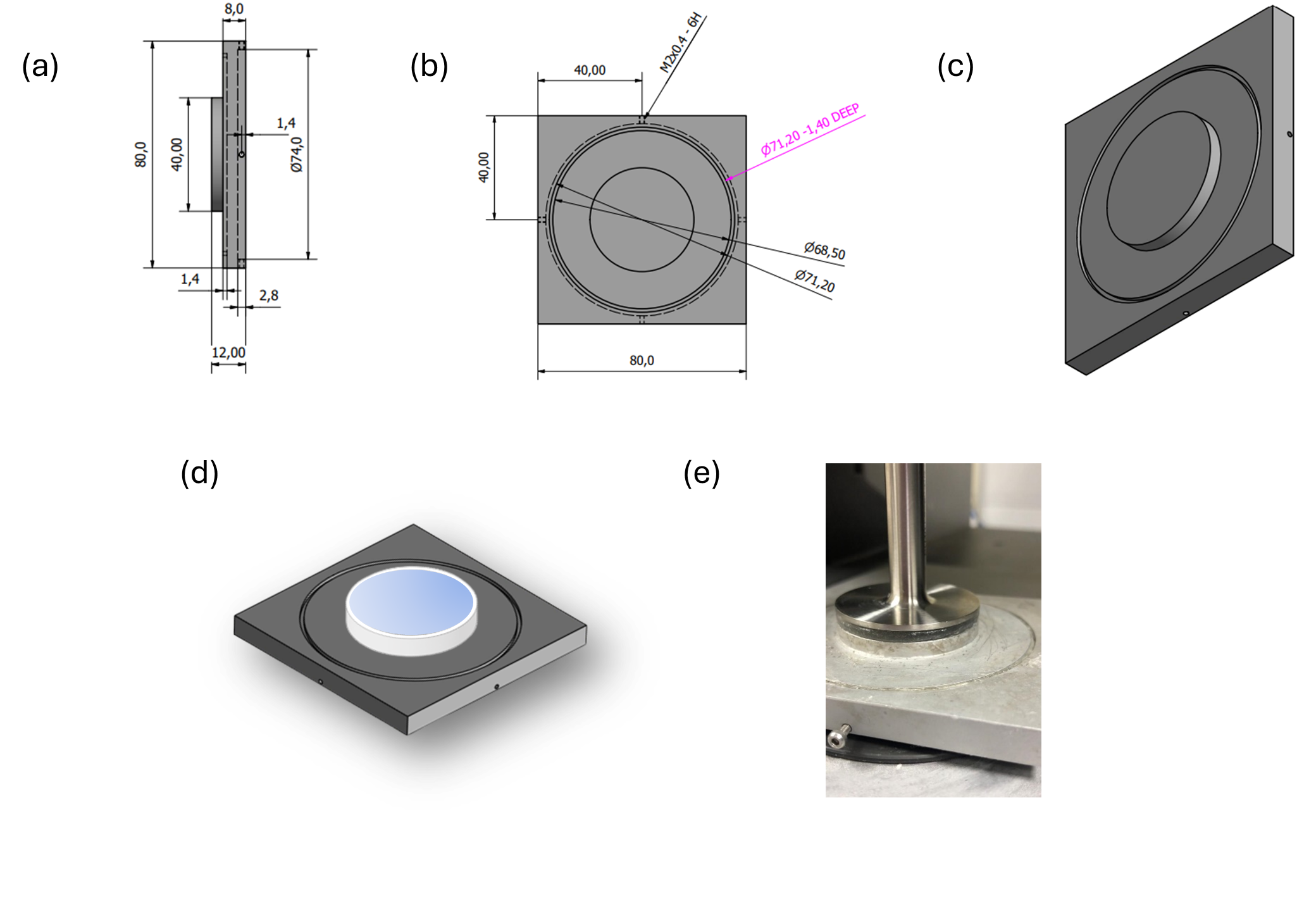}
\caption{\label{figure:SF1}Custom-designed mould. (a) Dimensions of the mould from the side view. Hydrogels sit on the raised-up portion with 40mm diameter and 4mm height; (b) Dimensions of the mould from the top view. Mould had 4 slots for screws that hold the mould tightly on the bottom plate of the rheometer; (c) Schematic of the mould. The top surface of the raised part was sandblasted; (d) A Teflon sleeve with an inner diameter of 40mm was inserted around the raised metallic part of the mould to contain the liquid reaction mixture. The height of the sleeve was 6mm which meant that the height of the hydrogels was 2mm; (e) Mould under the rheometer top plate during testing  
}
\end{figure*}
\newpage
\section{Investigating the effect of sandpaper}
\begin{figure}[!htb]
\includegraphics[width=0.9\textwidth]{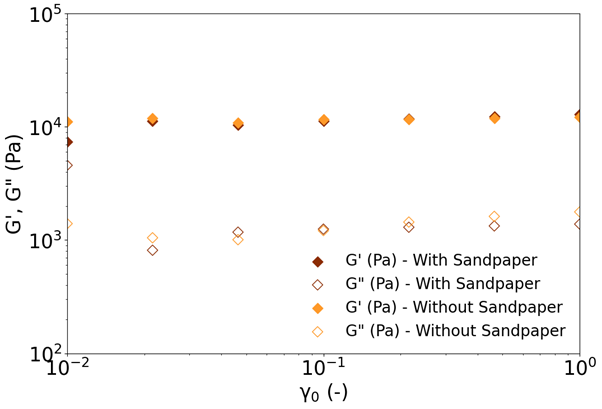}
\caption{\label{figure:SF2} Comparing frequency spectra of Alg+/PAAm+ hydrogel tested without sandpaper to with sandpaper. No considerable difference in frequency sweep was noted. 
}
\end{figure}
\newpage

\section{Zero amplitude tangent slope $G\rq_M$ and large amplitude secant slope $G\rq_L$}
\begin{figure}[!htb]
\includegraphics[width=0.9\textwidth]{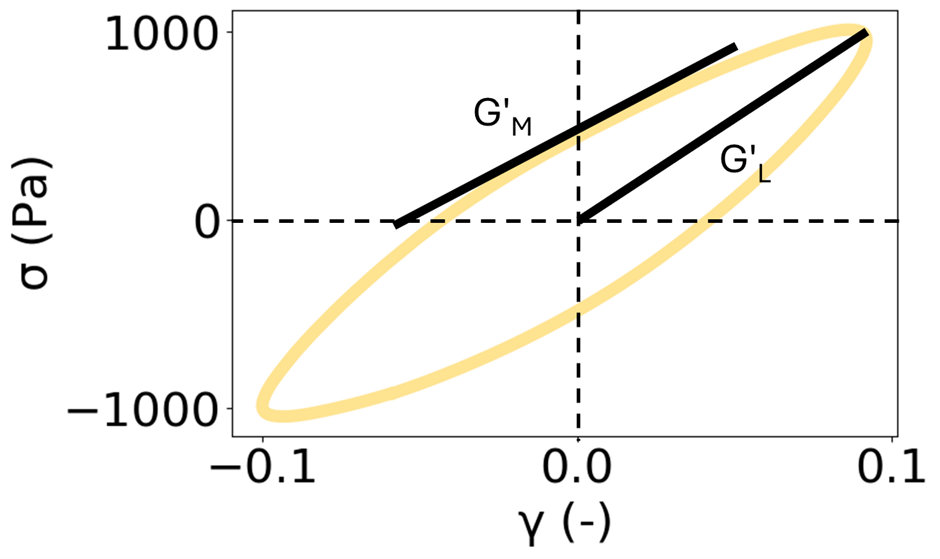}
\caption{\label{figure:SF3}Demonstration of zero amplitude tangent slope $G\rq_M$ and large amplitude secant slope $G\rq_L$.  
}
\end{figure}
\newpage

\section{Elastic stress evolution with oscillatory cycles increasing amplitude}
\begin{figure}[!htb]
\includegraphics[width=0.6\textwidth]{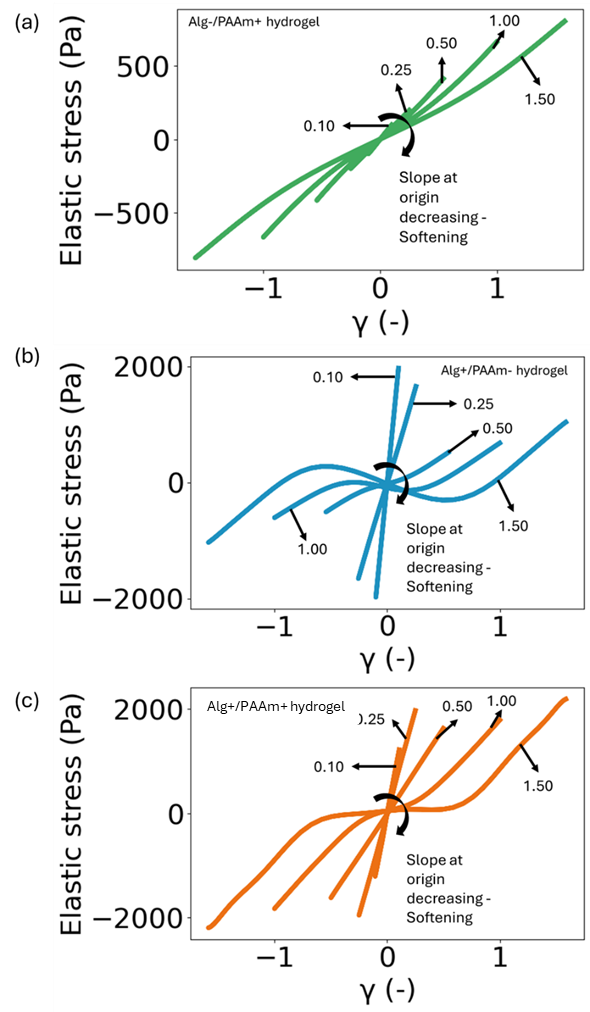}
\caption{\label{figure:SF4}Elastic stress evolution of hydrogels. (a) Alg-/PAAm+ hydrogel (b) Alg+/PAAm- hydrogel (c) Alg+/PAAm+ hydrogel. All hydrogels exhibit softening as the slope of elastic stress vs strain decreases with increasing amplitude of oscillatory cycles.  
}
\end{figure}
\newpage

\section{Evolution of Total Stress and its elastic and viscous stress components}
\begin{figure}[!htb]
\includegraphics[width=0.55\textwidth]{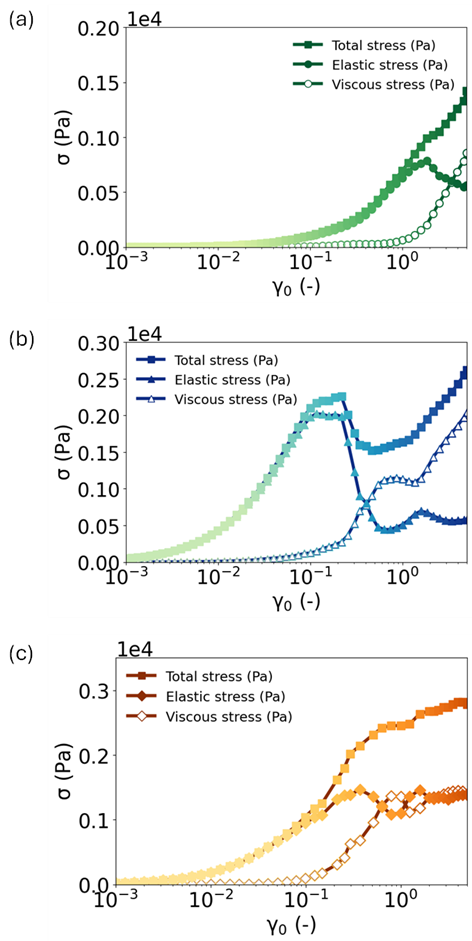}
\caption{\label{figure:SF5}Evolution of total stress and its elastic and viscous stress components. (a) Alg-/PAAm+ hydrogel (b) Alg+/PAAm- hydrogel (c) Alg+/PAAm+ hydrogel.  
}
\end{figure}
\newpage
\section{Comparison between elastic modulus and cage modulus in LVE regime}
\begin{figure}[!htb]
\includegraphics[width=0.9\textwidth]{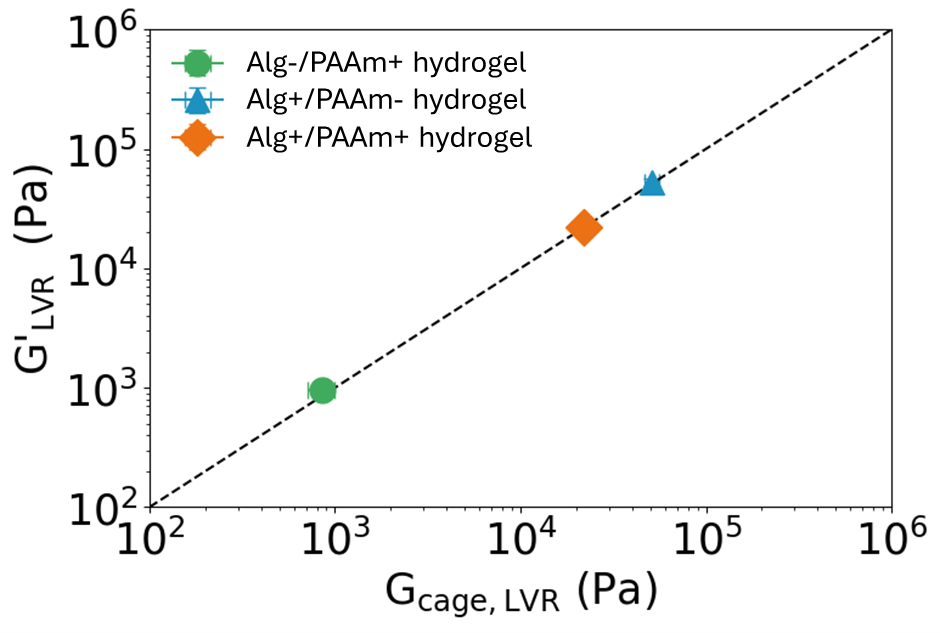}
\caption{\label{figure:SF6}Comparison between storage modulus and cage modulus in LVE regime. (a) Alg-/PAAm+ hydrogel (b) Alg+/PAAm- hydrogel (c) Alg/PAAm (Alg+/PAAm+) hydrogel. The representative dashed line represents the line on which elastic modulus and cage modulus will be equal. 
}
\end{figure}
\newpage
\section{Comparison of elastic stress contribution at the first peak for different agents}
\begin{figure}[!htb]
\includegraphics[width=0.8\textwidth]{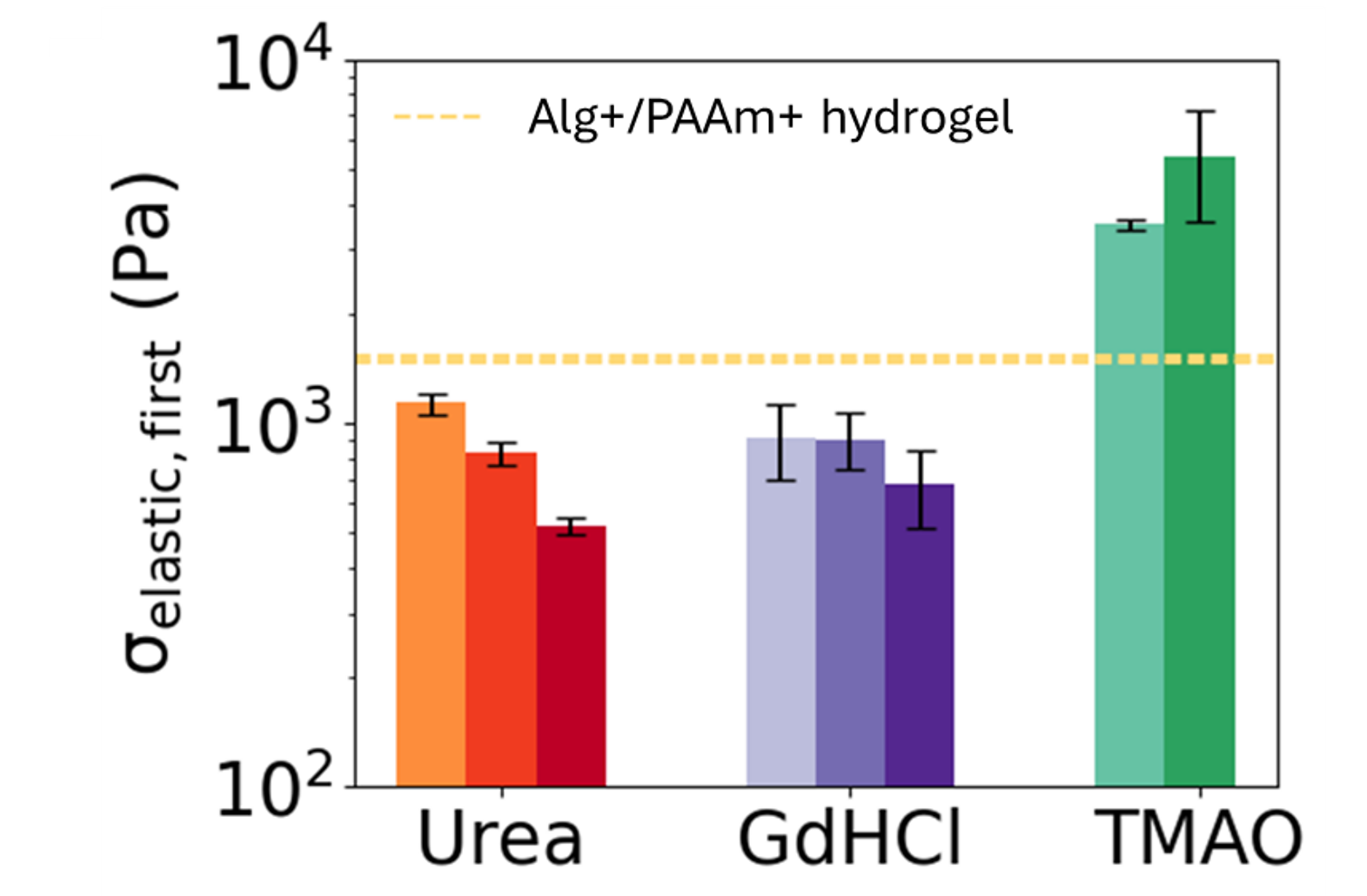}
\caption{\label{figure:SF7}Comparison of elastic stress contribution at the first peak for various hydrogen bond strength tuning agents. The figure shows that the reduction in the hydrogen bonds results in dip in the additional peak of elastic stress contribution for urea and GdHCl. The effect is reversed when the hydrogen bonds are strengthened by addition of TMAO. 
}
\end{figure}

\renewcommand{\thesection}{S\arabic{section}}  
\renewcommand{\thetable}{S\arabic{table}}  
\renewcommand{\thefigure}{S\arabic{figure}} 
\setcounter{figure}{0}   
\end{document}